\documentclass[1p]{elsarticle}

\journal{Physica A}

\begin{document}

\begin{frontmatter}

\title{The fitness landscapes of translation}

\author{Mario Josupeit and Joachim Krug}
\address{Institute for Biological Physics, University of Cologne,
  Z\"ulpicher Strasse 77, 50733 K\"oln, Germany}

\begin{abstract}
  Motivated by recent experiments on an antibiotic resistance gene,  
  we investigate genetic interactions between synonymous mutations in the framework of exclusion models of translation. We show
  that the range of possible interactions is markedly different depending on whether translation efficiency is assumed to
  be proportional to ribosome current or ribosome speed. In the first case every mutational effect has a definite sign that
  is independent of genetic background, whereas in the second case the effect-sign can vary depending on the presence of other
  mutations. The latter result is demonstrated using configurations of multiple translational bottlenecks induced by slow codons. 
\end{abstract}


\end{frontmatter}


\section{Introduction}

All living cells synthesize proteins by transcribing the hereditary information in their DNA into strands of messenger RNA
(mRNA) which
are subsequently translated into amino acid sequences. The genetic code that assigns triplets of nucleotides (codons) to
their corresponding amino acids is redundant, since most amino acids are encoded by several codons. Mutations that change
the DNA sequence (and hence the sequence of codons) but leave the amino acid sequence unchanged are called synonymous.
Such mutations were long thought to have no phenotypic consequences and hence to be evolutionarily silent.
However, meanwhile many cases have been reported where synonymous mutations profoundly affect organismal functions, primarily
by modifying the efficiency, timing and quality of protein production
\cite{Plotkin2011,Agashe2012,Hunt2014,Rodnina2016,Lebeuf-Taylor2019}.

Our work was motivated by a recent experimental study of synonymous mutations in a bacterial antibiotic resistance enzyme
that inactivates a class of drugs known as $\beta$-lactams. A panel of 10 synonymous point mutations was identified
which significantly increase the resistance against cefotaxime, as quantified by the drug concentration at which a fraction of
$10^{-4}$ of bacterial colonies survives \cite{Schenk2012}. Moreover, several of these
mutations display strong nonlinear interactions in their effect on resistance \cite{Zwart2018}. Of particular interest are
interactions referred to as sign-epistatic \cite{Weinreich2005}. In the present context
this implies that a mutation that increases resistance in the genetic background
of the original enzyme, and hence has a beneficial effect on bacterial survival, becomes deleterious in the presence
of another mutation, or vice versa. Sign epistasis is a key determinant of the structure of the fitness landscape
of an evolving population \cite{deVisser2014}.

Here we explore possible mechanisms that could explain sign epistatic interactions between synonymous mutations.
In line with the experimental observations \cite{Zwart2018}, we assume that the effect of the synonymous mutations
on organismal fitness is mediated by the efficiency of protein translation. The process of translation can be
described by stochastic particle models of exclusion type, which have been
used in the field for more than 50 years \cite{MacDonald1968,Zia2011,vanderHaar2012,Zur2016,Erdmann2020,Szavits2020a}.
These models treat
ribosomes as particles moving unidirectionally along a one-dimensional lattice of sites representing the codons.
The exclusion interaction ensuring that two particles cannot occupy the same site accounts for ribosome
queuing \cite{Szavits2020a,Diament2018}.

Within this conceptual framework we argue that the occurrence of sign epistatic interactions depends
crucially on the definition of translation efficiency. If efficiency is identified with the stationary ribosome
current, we prove rigorously that sign epistasis is not possible. In contrast, if efficiency is taken to be the
average speed of a ribosome (equivalently, the inverse of the ribosome travel time), then plausible scenarios
giving rise to sign-epistatic interactions are readily found, and fitness landscapes that are qualitatively similar
to those observed experimentally in \cite{Zwart2018} can be constructed \cite{Josupeit2020}.
We conclude, therefore, that there are (at least)
two different fitness landscapes of translation, one of which is always simple, whereas the other displays a complex
structure of hierarchically organized neutral networks.

\section{Inhomogeneous TASEP}

\begin{figure}[htb]
\centering
\includegraphics[width=0.7\linewidth]{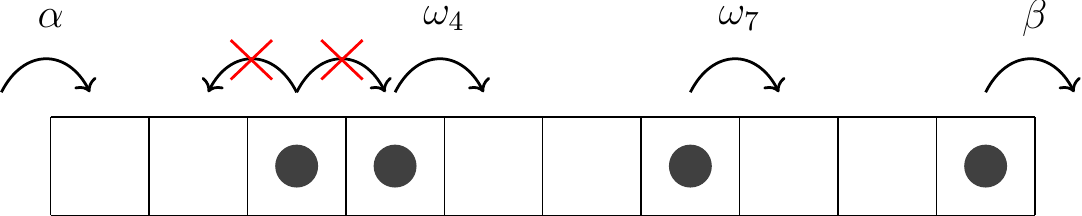}
\caption{\label{fig1} Schematic of the inhomogeneous TASEP on a lattice of $L=10$ sites. Particles enter the system at
  rate $\alpha$, hop from site $i$ to site $i+1$ at rate $\omega_i$ and exit at rate $\beta$. Backward jumps and
  jumps to occupied sites are forbidden. 
}
\end{figure}

The simplest model of translation is the totally asymmetric simple
exclusion process (TASEP) illustrated in Fig.~\ref{fig1}. The mRNA
strand is represented by a one-dimensional lattice of length $L$ and
ribosomes are particles occupying single lattice sites. Particles
enter the lattice at the \textit{initiation rate} $\alpha$ and leave
the lattice at the \textit{termination rate} $\beta$. Within the
lattice particles move from site $i$ to site $i+1$ at the
\textit{elongation rate} $\omega_i$ with $1 \leq i \leq L-1$.
This version of the TASEP neglects a number of important features of translation, such as the spatial extension of ribosomes and the
stepping cycle of the elongation process, which have been
included in more refined variants of the model
\cite{Dong2007,Ciandrini2010,Rudorf2015,Szavits2018b}.
Here we restrict ourselves to the simplest setting, as we
expect the conceptual points that we wish to make to be robust with
respect to such modifications. 

Importantly, the elongation rate $\omega_i$ depends on the identity of
the codon associated with the site $i$, which implies that the TASEP
is inhomogeneous with sitewise disorder \cite{Krug2000}. Whereas the
stationary state of the homogeneous TASEP with constant elongation
rates $\omega_i \equiv \omega$ is known exactly
\cite{Derrida1993,Schuetz1993}, only approximate approaches are
available for the inhomogeneous model \cite{Chou2004,Greulich2008,Szavits2013,Szavits2018a}. Key
observables of interest in the present context are the stationary
particle current $J$, which is site-independent because of mass
conservation, the site-dependent mean occupation numbers $\rho_i$, and
the spatially averaged mean particle density
\begin{equation}
  \label{rhobar}
  \bar{\rho} = \frac{1}{L}\sum_{i=1}^L \rho_i.
\end{equation}
Additionally we consider the mean travel time required for a particle
to move across the lattice, which is given by \cite{Szavits2020}
\begin{equation}
  \label{traveltime}
  T = \frac{L \bar{\rho}}{J} = \frac{L}{v} = L \tau. 
  \end{equation}
Here $v = J/\bar{\rho}$ denotes the average particle speed and $\tau = \frac{1}{v}$ the mean elongation time per site.

\section{Translation efficiency}

In order to quantify the effects of synonymous mutations on protein expression and fitness, we need to link the TASEP observables
defined above to the efficiency of translation. The most commonly used efficiency measure is the rate of protein production per
mRNA transcript, which (assuming steady-state conditions) corresponds to the stationary particle current $J$ \cite{Ciandrini2013}.
In experiments this quantity can be estimated as the ratio between the cellular abundance of a protein and that of the
corresponding mRNA \cite{Tuller2007}. In ribosome profiling experiments, which take snapshots of the ribosome occupancy along
the transcript, efficiency is instead associated with the fraction of codons that are covered by ribosomes, corresponding to
the mean particle density $\bar{\rho}$ in the TASEP \cite{Ingolia2009}. In the absence of ribosome queuing these two measures would be
expected to be proportional to each other, but empirically they are found to correlate poorly \cite{Plotkin2011},
indicating that queuing cannot generally be ignored.

The efficiency measures defined so far assume implicitly that ribosomes are sufficiently abundant, such that translation initiation
occurs readily as soon as the initiation site is free. However, the production of ribosomes is very costly for the cell, and
it has therefore been suggested that translation is optimized towards using each ribosome as efficiently
as possible \cite{Bulmer1991,Shah2011,Klumpp2012,Kavcic2020}.
Within the TASEP setting this implies that the relevant quantity associated
with translation efficiency is the ribosome travel time $T$ as a measure of translation cost \cite{Tuller2010,Mitarai2013}.
Equivalently, efficiency is quantified by the average speed $v$ of the ribosome.
In the following we examine the response of the ribosome current $J$ and the ribosome speed $v$ to synonymous mutations,
and argue that $v$ can display sign-epistatic interactions whereas $J$ cannot. 

\subsection{Ribosome current}

The stationary current in the inhomogeneous TASEP is a (generally unknown) function $J(\alpha,\beta,\omega_1,\dots,
\omega_{L-1})$ of the rates. In the Appendix we establish rigorously that this function is monotonic in its arguments. This implies
that a synonymous mutation that replaces one of the elongation rates $\omega_i$ by a rate $\tilde{\omega}_i$ always increases
the ribosome current if $\tilde{\omega}_i > \omega_i$ and decreases it if $\tilde{\omega}_i < \omega_i$, irrespective of the
values of the other rates. As a consequence, the effect-sign of any synonymous mutation is independent of the genetic background,
and sign epistasis cannot occur.


A numerical study of the inhomogeneous TASEP reported results for the stationary current that appear to contradict the
monotonicity property \cite{Foulaadvand2007}. The authors considered a binary system where a fraction $f$ of the sites
are assigned a slow rate $\omega_i = p_1 < 1$ and the remaining fraction $1-f$ have rate $\omega_i = 1$.
For certain conditions they observed that the stationary current was increasing with
increasing $f$, or decreasing with increasing $p_1$. We have repeated these simulations and find the expected monotonic
behavior throughout \cite{Josupeit2020}. The results reported in \cite{Foulaadvand2007} may have been caused by insufficient
relaxation to the stationary state\footnote{M.E. Foulaadvand, private communication.}.

\subsection{Ribosome speed in a bottleneck configuration}

We base the discussion of the ribosome speed on a simple rate configuration with a single slow site (the `bottleneck')
with rate $\omega_k = b < 1$ embedded in a homogeneous TASEP with rates $\omega_i = 1$ for $i \neq k$. The TASEP with a
bottleneck has been the subject of numerous studies \cite{Janowsky1994,Greulich2008a,Schmidt2015},
and although its stationary state is not known exactly, many features of the model are well understood. For convenience we
focus on the case where the initiation and termination rates are large, $\alpha = \beta = 1$, such that the behavior of
the system is dominated by the bottleneck. In the limit of large $L$ the system then phase separates into a high density
region (the `traffic jam') preceding the bottleneck and a low density region behind it. Continuity of the current implies
that the densities of the two regions are related by $\varrho \equiv \rho_\mathrm{low} = 1 - \rho_\mathrm{high}$. The
density $\varrho$ is determined through the relation $\varrho(1-\varrho) = j(b)$, where $j(b)$ denotes the maximal
current that can flow through the bottleneck site. The function $j(b)$ is not explicitly known, but it has been established
that $j(b) < j(1) = \frac{1}{4}$ for any $b < 1$ \cite{Basu2014}. A simple mean field argument yields the approximate expression
$j(b) = \frac{b}{(1+b)^2}$, and more accurate power series expansions can be found in \cite{Szavits2013} and
\cite{Janowsky1994}. Importantly, $\varrho \in (0,\frac{1}{2})$ is uniquely determined by $b$, and we may therefore
use $\varrho$ to quantify the strength of the bottleneck.

Whereas the current is fully determined by the bottleneck strength, the ribosome travel time and speed also depend
on its location $k$. As we are generally concerned with the limit of large $L$, it is useful to introduce
the scaled position $x = k/L$ for $k,L \to \infty$. Then the average density in the stationary state is given by
\begin{equation}
  \label{bottle1}
  \bar{\rho}(\varrho,x) = x (1-\varrho) + (1-x) \varrho
\end{equation}
and correspondingly the mean elongation time is
\begin{equation}
  \label{bottle2}
  \tau = \frac{\bar{\rho}}{\varrho(1-\varrho)} = \frac{x}{\varrho} + \frac{1-x}{1- \varrho} =
  \frac{1}{1-\varrho} + x \frac{1 - 2 \varrho}{\varrho(1-\varrho)}.
  \end{equation}
Since $\varrho < \frac{1}{2}$, the elongation time is an increasing function of $x$. It is straightforward to show that
for $x < \varrho$ the elongation time is smaller than the value $\tau = 2$ in the absence of a bottleneck.
Thus by placing the bottleneck sufficiently close to the initiation site the ribosome travel time can actually be
reduced compared to the homogeneous system \cite{Mitarai2013}. This effect has been invoked to explain the evolutionary
benefits of the accumulation of slow codons at the beginning of mRNA transcripts that is observed in genomic data
\cite{Tuller2010}.

\begin{figure}[htb]
\centering
\includegraphics[width=0.8\linewidth]{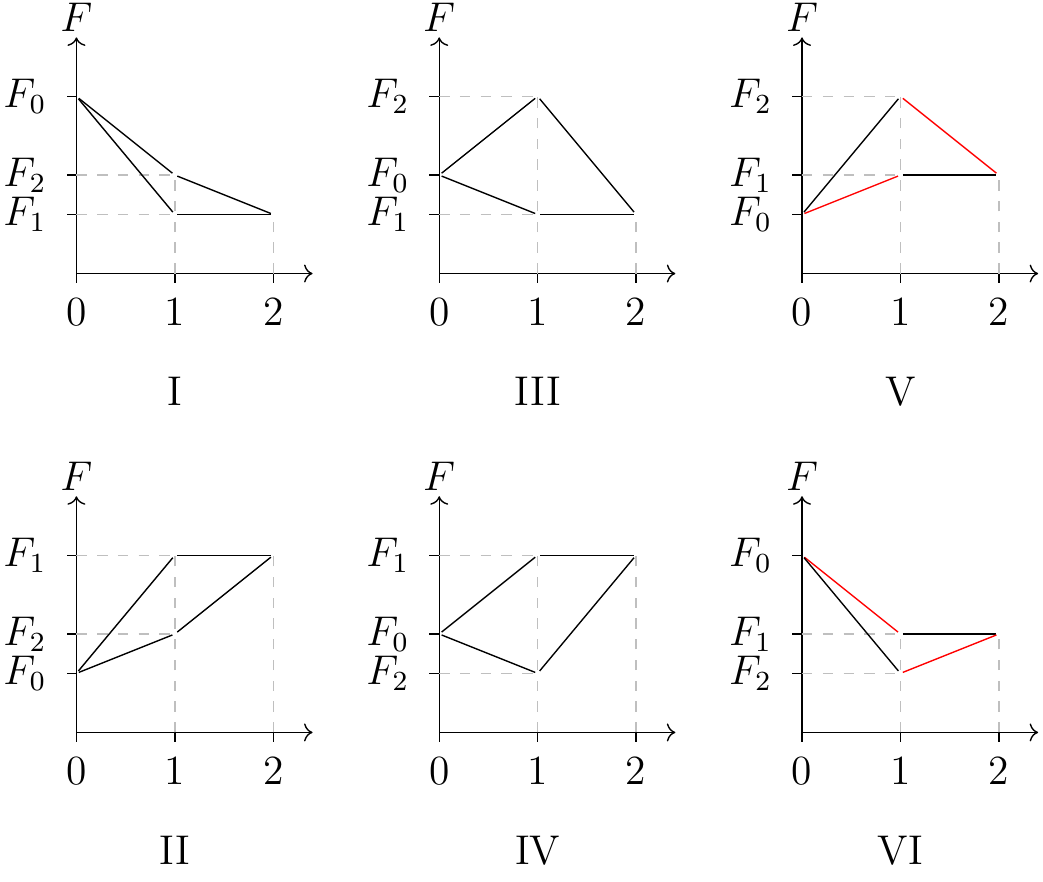}
\caption{\label{fig3} All possible fitness landscapes for a system with two bottlenecks. The plots show the fitness values
  of the four genotypes as a function of the number of mutations. In all cases the fitness of the double mutant is
  equal to $F_1$. If fitness is associated with the ribosome current the landscapes has to be of type I. Landscapes
  V and VI display instances of sign epistasis which are marked in red.
}
\end{figure}

\section{Multiple bottlenecks}

Consider now a situation where synonymous point mutations at two sites $k_1$ and $k_2$ of the mRNA
replace regular codons with elongation rate $\omega_i = 1$ by bottlenecks with rates $b_1, b_2 < 1$. The scaled positions
of the two sites are denoted by $x_\nu = k_\nu/L$, $\nu = 1,2$, and we assume without loss of generality that $b_1 < b_2$.
This defines a simple genetic system comprising four genotypes, the unmutated sequence $(0)$, the single mutants $\nu = 1,2$ and
the double mutant (12) in which both bottlenecks are present. Since the maximal currents supported by the bottlenecks satisfy
$j(b_1) < j(b_2)$, bottleneck 2 is irrelevant in the presence of bottleneck 1. As a consequence, the double mutant
is phenotypically indistinguishable from the system containing only bottleneck 1. We say that mutation 2 is
\textit{conditionally neutral} in the presence of mutation 1. 

\begin{figure}[htb]
\centering
\includegraphics[width=\linewidth]{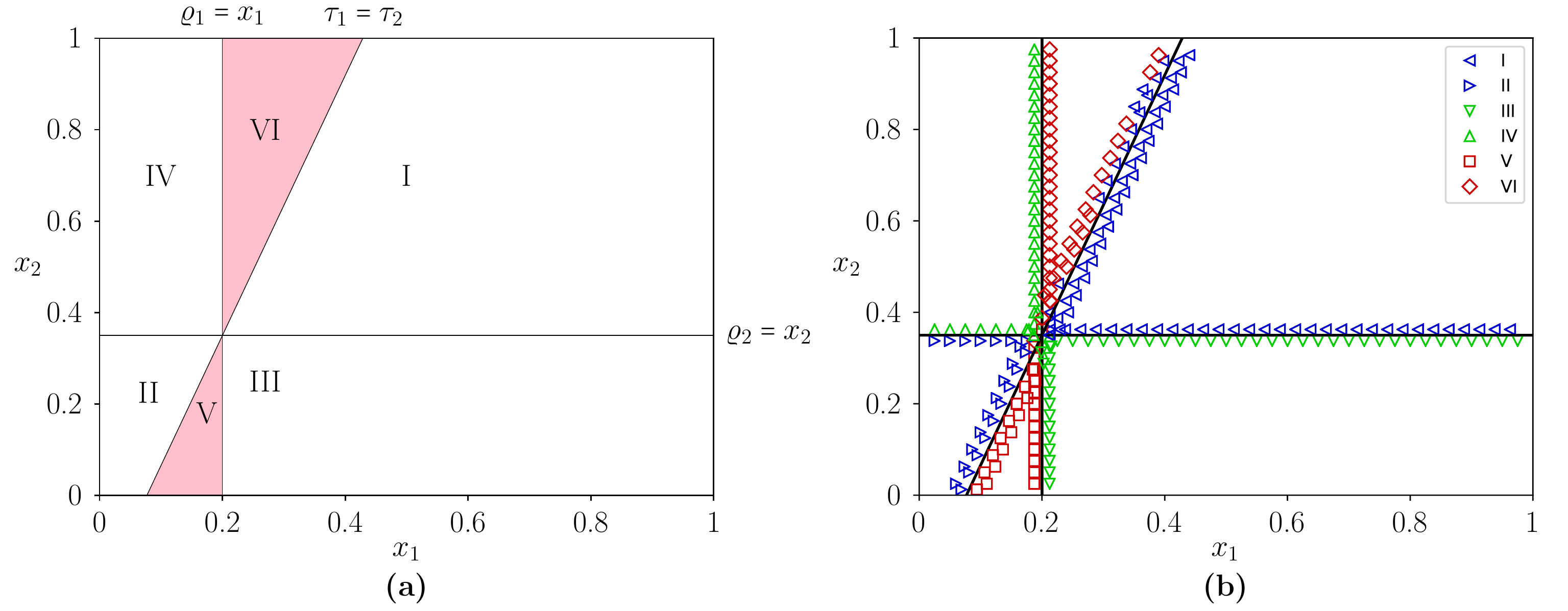}
\caption{\label{fig4} (a) Predicted phase diagram in the plane of scaled bottleneck positions $(x_1,x_2)$ displaying the regions where
  the orderings I-VI depicted in Fig.~\ref{fig3} are realized for the ribosome speed.
  Along the vertical line $x_1 = \varrho_1$ the travel time of the
  system with bottleneck 1 is equal to the travel time of the system without bottlenecks, $\tau_1 = \tau_0 = 2$,
  and correspondingly along the horizontal line $x_2 = \varrho_2$ the condition $\tau_2 = 2$ holds. The slanted line is determined by the condition $\tau_1 = \tau_2$. In the shaded regions V and VI sign epistasis is present.
  (b) Numerical verification of the predicted phase diagram based on simulations of a TASEP with $L=800$. The stationary current
  $J$ and density $\bar{\rho}$ were determined for the two bottlenecked systems by averaging over $5 \times 10^6$ time steps
  following an equilibration period of $10^5$ time steps. The travel times were computed from (\ref{traveltime}),
  and compared to the travel time $\tau_0$ of the system without bottlenecks obtained from a separate simulation. Symbols
  show the ordering of the simulated travel times $\tau_0$, $\tau_1$ and $\tau_2$.
  The bottleneck densities chosen for this image are $\varrho_1 = 0.2$ and $\varrho_2 = 0.35$.
}
\end{figure}

The fitness landscape spanned by the four genotypes thus contains 3 distinct fitness levels that we denote by $F_0$, $F_1 = F_{12}$
and $F_2$. If fitness is associated with the ribosome current the ordering of the fitness values is dictated by the strength
of the bottlenecks and given by $F_0 > F_2 > F_1 = F_{12}$. By contrast, through a suitable choice of the bottleneck positions,
for the elongation time or the ribosome speed all $3! = 6$ possible orderings of fitness values can be realized. The resulting
fitness landscapes are illustrated in Fig.~\ref{fig3}. In 2 of the 6 cases sign epistasis occurs, in that the addition of
bottleneck 1 can either increase or decrease the ribosome speed depending on the presence of bottleneck 2.
Using the expression (\ref{bottle2}) for the elongation time, the different orderings can be mapped to regions in the
plane of scaled bottleneck positions $(x_1,x_2) \in [0,1]^2$. Because of the linearity of (\ref{bottle2}) these regions
are delimited by straight lines (Fig.~\ref{fig4}). Figure \ref{fig4}(b) shows a phase diagram obtained
from simulations of finite systems, which agrees very well with the
prediction based on the asymptotic behavior for $L \to \infty$.

\begin{figure}[htb]
\centering
\includegraphics[width=0.8\linewidth]{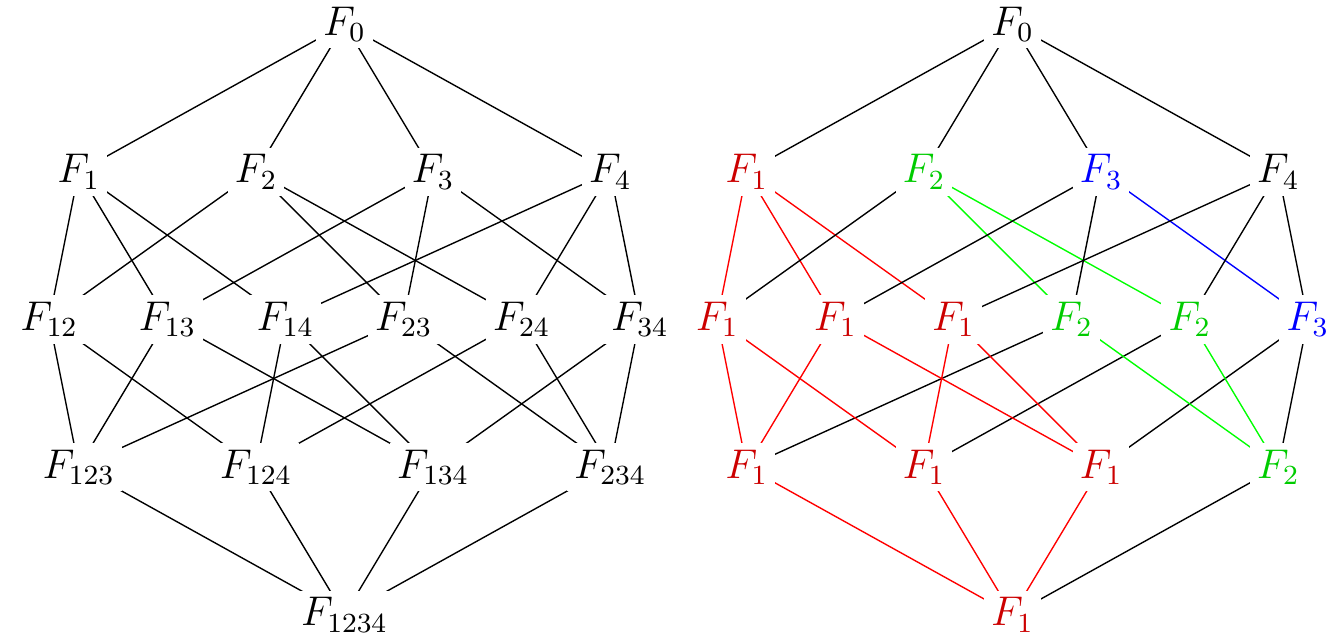}
\caption{\label{fig5} (a) The $2^4 = 16$ genotypes in a system with 4 bottlenecks form a 4-dimensional hypercube, where neighboring
  nodes are connected by the addition or removal of a bottleneck. The indices of the fitness values show which bottlenecks
  are present in the genotype. (b) Under the model the hybercube decomposes into $N-1$ subcubes
  containing $2^K$ genotypes each ($K = N-1, N-2, \dots, 1$) on which the fitness is constant and determined by the strongest
  bottleneck that is present. In addition there are two single nodes corresponding to the system without a bottleneck ($F_0$)
  and the weakest bottleneck ($F_N$). In the figure the subcubes of size 8, 4 and 2 are marked in color. 
}
\end{figure}

These considerations generalize straightforwardly to systems with $N > 2$ potential bottlenecks. A genotype is specificied by
the presence or absence of each of the bottlenecks, and correspondingly there are $2^N$ genotypes in total. The phenotype
(ribosome current or ribosome speed) of a given genotype is determined by the strongest bottleneck that is present in the
system. Labeling the bottlenecks in decreasing order of their strengths, $b_1 < b_2 < \dots < b_N < 1$, it follows that
all $2^{N-1}$ genotypes in which bottleneck 1 is present share the same fitness value $F_1$. Among the $2^{N-1}$ genotypes
that lack bottleneck $1$, half are dominated by bottleneck 2, and so on. This leads to a hierarchical structure of neutral
regions that is illustrated in Fig.~\ref{fig5} for $N=4$. If fitness is taken to be proportional to the ribosome
current, the $N+1$ fitness values are ordered as $F_1 < F_2 < \dots < F_N < F_0$, whereas all possible $(N+1)!$ orderings
can be realized for the ribosome speed.

\section{Conclusion}

In this article we have outlined a possible scenario for the occurrence of sign-epistatic interactions between the effects of
synonymous mutations on the efficiency of protein production. A necessary requirement for the scenario to apply is that
translation efficiency is related to ribosome speed rather than to ribosome current. We have worked out a detailed quantitative
description for the idealized situation of a mRNA transcript with homogeneous elongation rates into which a small number of
well-separated slow codons (``bottlenecks'') are inserted, but we expect our considerations to generalize to more realistic
elongation rate profiles. The key feature of the translation process that enables sign epistasis is the
fact that, somewhat counter-intuitively, the introduction of slow codons can decrease the ribosome travel time by reducing
the amount of ribosome queuing \cite{Kavcic2020,Tuller2010,Mitarai2013}.
To the extent that ribosome queuing is a generic feature of translation \cite{Szavits2020a,Diament2018},
sign-epistatic interactions in the ribosome speed are therefore likely to arise.

The model of discrete, well-separated bottlenecks predicts a specific fitness landscape
structure where the hypercube of genotypes decomposes into lower-dimensional subcubes of constant fitness (Fig.~\ref{fig5}).
The combination of strong sign-epistatic interactions with extended plateaux of approximately constant resistance levels
is indeed a visually striking feature of the experimental data set that motivated this work \cite{Zwart2018}. A quantitative
comparison between the model and the data is however beyond the scope of this article and will be presented elsewhere.

\section*{Acknowledgments} JK is grateful to Martin Evans, Grzegorz Kudla, Mamen Romano and
Juraj Szavits-Nossan for helpful discussions, and to SUPA and the Higgs Centre for Theoretical Physics in Edinburgh
for their gracious
hospitality during the early stages of the project. The work was supported by DFG within CRC 1310 \textit{Predictability
in Evolution}. We dedicate this article to the memory of
Dietrich Stauffer, fearless explorer of disciplinary boundaries and
translator of scientific and cultural idioms.   

\section*{Appendix: Proof of the monotonicity property}

We prove that the stationary current in the inhomogeneous TASEP with open
boundaries is a monotonic function of the jump rates.
The proof is based on the waiting time representation of the TASEP explained in
\cite{Krug1994}. Briefly, the TASEP occupation variables $\sigma_i \in \{0,1\}$, $i=1,\dots,L$, are mapped to a single-step
(SS) interface configuration $h(i)$ through the relation
\begin{equation}
  \label{SSheight}
  \sigma_i = \frac{1}{2}\left[1 + h(i) - h(i+1) \right]
\end{equation}
such that a growth event $h(i) \to h(i)+2$ corresponds to the jump of a particle from
site $i-1$ to site $i$. Thus the height variables measure the time-integrated local particle
current. A dual description of the SS growth process is provided by the waiting time
variables $t(i,j)$ denoting the time at which the interface height reaches the point $(i,j)$
on the underlying tilted square lattice, i.e. when $h(i) = j$ (see Fig.~1 of \cite{Krug1994} for an illustration
of the geometry). The SS/TASEP growth rule implies that
\begin{equation}
  \label{recursion}
  t(i,j) = \eta(i,j) + \max \left[t(i-1,j-1), t(i+1,j-1) \right],
\end{equation}
where the $\eta(i,j)$ are exponentially distributed random variables.
For the open boundary TASEP on a lattice of $L$ sites the recursion (\ref{recursion}) holds for $2 \leq i \leq L-1$
with some modifications at the boundary sites $i=1$ and $i=L$. These modifications
are not important here as the initiation and termination rates $\alpha$ and $\beta$ will be assumed to be fixed.

The solution of (\ref{recursion}) can be expressed as
\begin{equation}
  \label{solution}
  t(i,j) = \max_{\pi \in P_{i,j}} {\cal{T}}(\pi)
\end{equation}
where $P_{i,j}$ is the set of upward directed paths $\pi$ on the tilted square lattice that end at
$(i,j)$, and the passage time ${\cal{T}}(\pi)$ of a path is the sum of the random variables along
the path,
\begin{equation}
  \label{T}
{\cal{T}}(\pi) = \sum_{(x,y) \in \pi} \eta(x,y).
\end{equation}
The problem of finding the solution of (\ref{solution}) is known as last
passage percolation \cite{Basu2014,Krug1991}.

At long times the SS interface attains some stationary velocity $c > 0$ such that
$\lim_{t \to \infty} \frac{h(i)}{t} = c$ independent of $i$. Since the height increases by 2 for each jump of
a TASEP particle, the velocity is related to the stationary current $J$ through $c = 2J$.
Correspondingly, by the law of large numbers the rescaled waiting times converge as
\begin{equation}
  \label{limit}
  \lim_{h \to \infty} \frac{t(i,h)}{h} = \frac{1}{c} = \frac{1}{2J}
\end{equation}
which expresses the stationary current in terms of the last passage percolation problem (\ref{solution}).

In the inhomogeneous TASEP the distribution of the random waiting times $\eta(i,j)$
depends on the site label $i$. Specifically,
the probability density of $\eta(i,j)$ is given by
$p_i(\eta) = \omega_i e^{-\omega_i \eta}$.
  To generate a realization of the TASEP process for a given set of rates $\{\omega_i\}_{i=1,\dots,L-1}$,
  we first draw a set of uniform random variables $u(i,j) \in [0,1]$ which are subsequently converted
  to exponentially distributed waiting times according to
  \begin{equation}
   \label{u}
\eta(i,j)  = -\frac{1}{\omega_i} \ln u(i,j).
\end{equation}
Fixing the uniform random variables, we may now compare two inhomogeneous TASEP's
with jump rates $\{\omega_i\}$ and $\{\tilde{\omega}_i\}$, where $\omega_i \leq \tilde{\omega}_i$
for all $i$. It then follows from (\ref{u}) that $\eta(i,j) \geq \tilde{\eta}(i,j)$ for all $i,j$.
Correspondingly, (\ref{T}) and (\ref{solution})  imply that $t(i,j) \geq \tilde{t}(i,j)$ and
therefore $\tilde{J} \geq J$ according to (\ref{limit}).

\bibliography{mybibfile}

\begin{thebibliography}{10}
\expandafter\ifx\csname url\endcsname\relax
  \def\url#1{\texttt{#1}}\fi
\expandafter\ifx\csname urlprefix\endcsname\relax\def\urlprefix{URL }\fi
\expandafter\ifx\csname href\endcsname\relax
  \def\href#1#2{#2} \def\path#1{#1}\fi

\bibitem{Plotkin2011}
J.~B. Plotkin, G.~Kudla, Synonymous but not the same: the causes and
  consequences of codon bias, Nat. Rev. Genet. 12 (2011) 32.

\bibitem{Agashe2012}
D.~Agashe, N.~C. Martinez-Gomez, D.~A. Drummond, C.~J. Marx, Good codons, bad
  transcript: large reductions in gene expression and fitness arising from
  synonymous mutations in a key enzyme, Mol. Biol. Evol. 30 (2012) 549.

\bibitem{Hunt2014}
R.~C. Hunt, V.~L. Simhadri, M.~Iandoli, Z.~E. Sauna, C.~Kimchi-Sarfaty,
  Exposing synonymous mutations, Trends in Genetics 30 (2014) 308.

\bibitem{Rodnina2016}
M.~V. Rodnina, The ribosome in action: Tuning of translational efficiency and
  protein folding, Protein Science 25 (2016) 1390.

\bibitem{Lebeuf-Taylor2019}
E.~Lebeuf-Taylor, N.~McCloskey, S.~F. Bailey, A.~Hinz, R.~Kassen, The
  distribution of fitness effects among synonymous mutations in a gene under
  directional selection, eLife 8 (2019) e45952.

\bibitem{Schenk2012}
M.~F. Schenk, I.~G. Szendro, J.~Krug, J.~A. G.~M. de~Visser, Quantifying the
  adaptive potential of an antibiotic resistance enzyme, PloS Genetics 8 (2012)
  e1002783.

\bibitem{Zwart2018}
M.~P. Zwart, M.~F. Schenk, S.~Hwang, B.~Koopmanschap, N.~de~Lange, L.~van~der
  Pol, T.~T.~T. Nga, I.~G. Szendro, J.~Krug, J.~A. G.~M. de~Visser, Unraveling
  the causes of adaptive benefits of synonymous mutations in {TEM-1
  $\beta$-lactamase}, Heredity 121 (2018) 406.

\bibitem{Weinreich2005}
D.~M. Weinreich, R.~A. Watson, L.~Chao, Perspective: Sign epistasis and genetic
  constraint on evolutionary trajectories, Evolution 59 (2005) 1165.

\bibitem{deVisser2014}
J.~A. G.~M. de~Visser, J.~Krug, Empirical fitness landscapes and the
  predictability of evolution, Nat. Rev. Genet. 15 (2014) 480.

\bibitem{MacDonald1968}
C.~T. MacDonald, J.~H. Gibbs, A.~C. Pipkin, Kinetics of biopolymerization on
  nucleic acid templates, Biopolymers 6 (1968) 1.

\bibitem{Zia2011}
R.~K.~P. Zia, J.~J. Dong, B.~Schmittmann, Modeling translation in protein
  synthesis with {TASEP}: A tutorial and recent developments, J. Stat. Phys.
  144 (2011) 405.

\bibitem{vanderHaar2012}
T.~van~der Haar, Mathematical and computational modelling of ribosomal movement
  and protein synthesis: an overview, Computational and Structural
  Biotechnology Journal 1 (2012) e201204002.

\bibitem{Zur2016}
H.~Zur, T.~Tuller, Predictive biophysical modeling and understanding of the
  dynamics of {mRNA} translation and its evolution, Nucleic Acids Research 44
  (2016) 9031.

\bibitem{Erdmann2020}
D.~D. Erdmann-Pham, K.~D. Duc, Y.~S. Song, The key parameters that govern
  translation efficiency, Cell Systems 10 (2020) 183.

\bibitem{Szavits2020a}
J.~Szavits-Nossan, L.~Ciandrini, Inferring efficiency of translation initiation
  and elongation from ribosome profiling, Nucleic Acids Research (advance
  article).

\bibitem{Diament2018}
A.~Diament, A.~Feldman, E.~Schochet, M.~Kupiec, Y.~Arava, T.~Tuller, The extent
  of ribosome queuing in budding yeast, PLoS Comput. Biol. 14 (2018) e1005951.

\bibitem{Josupeit2020}
M.~Josupeit, The fitness landscape of translation, MSc thesis, University of
  Cologne, 2020.

\bibitem{Dong2007}
J.~J. Dong, B.~Schmittmann, R.~K.~P. Zia, Towards a model for protein
  production rates, J. Stat. Phys. 128 (2007) 21.

\bibitem{Ciandrini2010}
L.~Ciandrini, I.~Stansfield, M.~C. Romano, Role of the particle's stepping
  cycle in an asymmetric exclusion process: A model of {mRNA} translation,
  Phys. Rev. E 81 (2010) 051904.

\bibitem{Rudorf2015}
S.~Rudorf, R.~Lipowsky, Protein synthesis in {E. coli}: Dependence of
  codon-specific elongation on {tRNA} concentration and codon usage, PLoS ONE
  10 (2015) e0134994.

\bibitem{Szavits2018b}
J.~Szavits-Nossan, L.~Ciandrini, M.~C. Romano, Deciphering {mRNA} sequence
  determinants of protein production rate, Phys. Rev. Lett. 120 (2018) 128101.

\bibitem{Krug2000}
J.~Krug, Phase separation in disordered exclusion models, Braz. J. Phys. 30
  (2000) 97.

\bibitem{Derrida1993}
B.~Derrida, M.~R. Evans, V.~Hakim, V.~Pasquier, Exact solution of a {1D}
  asymmetric exclusion model using a matrix formulation, J. Phys. A 26 (1993)
  1493.

\bibitem{Schuetz1993}
G.~Sch{\"u}tz, E.~Domany, Phase transitions in an exactly soluble
  one-dimensional exclusion process, J. Stat. Phys. 72 (1993) 277.

\bibitem{Chou2004}
T.~Chou, G.~Lakatos, Clustered bottlenecks in {mRNA} translation and protein
  synthesis, Phys. Rev. Lett. 93 (2004) 198101.

\bibitem{Greulich2008}
P.~Greulich, A.~Schadschneider, Single-bottleneck approximation for driven
  lattice gases with disorder and open boundary conditions, J. Stat. Mech.:
  Theory Exp. 2008 (2008) P04009.

\bibitem{Szavits2013}
J.~Szavits-Nossan, Disordered exclusion process revisited: some exact results
  in the low-current regime, J. Phys. A 46 (2013) 315001.

\bibitem{Szavits2018a}
J.~Szavits-Nossan, M.~C. Romano, L.~Ciandrini, Power series solution of the
  inhomogeneous exclusion process, Phys. Rev. E 97 (2018) 052139.

\bibitem{Szavits2020}
J.~Szavits-Nossan, M.~R. Evans, Dynamics of ribosomes in {mRNA} translation
  under steady- and nonsteady-state conditions, Phys. Rev. E 101 (2020) 062404.

\bibitem{Ciandrini2013}
L.~Ciandrini, I.~Stansfield, M.~C. Romano, Ribosome traffic on {mRNAs} maps to
  gene ontology: {Genome-wide} quantification of translation initiation rates
  and polysome size regulation, PLoS Comp. Biol. 9 (2013) e1002866.

\bibitem{Tuller2007}
T.~Tuller, M.~Kupiec, E.~Ruppin, Determinants of protein abundance and
  translation efficiency in \textit{S. cerevisiae}, PLoS Comp. Biol. 3 (2007)
  248.

\bibitem{Ingolia2009}
N.~T. Ingolia, S.~Ghaemmaghami, J.~R.~S. Newman, J.~S. Weissman, Genome-wide
  analysis in vivo of translation with nucleotide resolution using ribosome
  profiling, Science 324 (2009) 218.

\bibitem{Bulmer1991}
M.~Bulmer, The mutation-selection-drift theory of synonymous codon usage,
  Genetics 129 (1991) 897.

\bibitem{Shah2011}
P.~Shah, M.~A. Gilchrist, Explaining complex codon usage patterns with
  selection for translational efficiency, mutation bias, and genetic drift,
  PNAS 108 (2011) 10231.

\bibitem{Klumpp2012}
S.~Klumpp, J.~Dong, T.~Hwa, On ribosome load, codon bias and protein abundance,
  PLoS ONE 7 (2012) e48542.

\bibitem{Kavcic2020}
B.~Kav\v{c}i\v{c}, G.~Tka\v{c}ik, T.~Bollenbach, Mechanisms of drug
  interactions between translation-inhibiting antibiotics, Nat. Commun. 11
  (2020) 4013.

\bibitem{Tuller2010}
T.~Tuller, A.~Carmi, K.~Vestsigian, S.~Navon, Y.~Dorfan, J.~Zaborske, T.~Pan,
  O.~Dahan, I.~Furman, Y.~Pilpel, An evolutionarily conserved mechanism for
  controlling the efficiency of protein translation, Cell 141 (2010) 344.

\bibitem{Mitarai2013}
N.~Mitarai, S.~Pedersen, Control of ribosome traffic by position-dependent
  choice of synonymous codons, Phys. Biol. 10 (2013) 056011.

\bibitem{Foulaadvand2007}
M.~E. Foulaadvand, S.~Chaaboki, M.~Saalehi, Characteristics of the asymmetric
  simple exclusion process in the presence of quenched spatial disorder, Phys.
  Rev. E 75 (2007) 011127.

\bibitem{Janowsky1994}
S.~A. Janowsky, J.~L. Lebowitz, Exact results for the asymmetric simple
  exclusion process with a blockage, J. Stat. Phys. 77 (1994) 35.

\bibitem{Greulich2008a}
P.~Greulich, A.~Schadschneider, Phase diagram and edge effects in the {ASEP}
  with bottlenecks, Physica A 387 (2008) 1972.

\bibitem{Schmidt2015}
J.~Schmidt, V.~Popkov, A.~Schadschneider, Defect-induced phase transition in
  the asymmetric simple exclusion process, EPL 110 (2015) 20008.

\bibitem{Basu2014}
R.~Basu, V.~Sidoravicius, A.~Sly, Last passage percolation with a defect line
  and the solution of the slow bond problem, Preprint arXiv (2014) 1408.3463.

\bibitem{Krug1994}
J.~Krug, L.-H. Tang, Disorder-induced unbinding in confined geometries, Phys.
  Rev. E 50 (1994) 104.

\bibitem{Krug1991}
J.~Krug, H.~Spohn, Kinetic roughening of growing surfaces, in: C.~Godr\`eche
  (Ed.), Solids Far From Equilibrium, Cambridge University Press, Cambridge,
  1991, p. 479.

\end{thebibliography}

\end{document}